\input lanlmac.tex

\overfullrule=0pt
\input epsf.tex
\figno=0
\def\fig#1#2#3{
\par\begingroup\parindent=0pt\leftskip=1cm\rightskip=1cm\parindent=0pt
\baselineskip=11pt
\global\advance\figno by 1
\midinsert
\epsfxsize=#3
\centerline{\epsfbox{#2}}
\vskip 12pt
{\bf Fig. \the\figno:} #1\par
\endinsert\endgroup\par
}
\def\figlabel#1{\xdef#1{\the\figno}}
\def\encadremath#1{\vbox{\hrule\hbox{\vrule\kern8pt\vbox{\kern8pt
\hbox{$\displaystyle #1$}\kern8pt}
\kern8pt\vrule}\hrule}}
 
%

\def\np#1#2#3{{\it Nucl. Phys.} {\bf B#1} (#2) #3}
\def\pl#1#2#3{{\it Phys. Lett. }{\bf B#1} (#2) #3}
\def\prl#1#2#3{{\it Phys. Rev. Lett.}{\bf #1} (#2) #3}
\def\physrev#1#2#3{{\it Phys. Rev.} {\bf D#1} (#2) #3}
\def\prb#1#2#3{{\it Phys. Rev.} {\bf B#1} (#2) #3}

\def\rmatp#1#2#3{{\it Rev. Math. Phys. }{\bf #1} (#2) #3}
\def\cmp#1#2#3{{\it Comm. Math. Phys.} {\bf #1} (#2) #3}
\def\mpl#1#2#3{{\it Mod. Phys. Lett. }{\bf #1} (#2) #3}
\def\ijmp#1#2#3{{\it Int. J. Mod. Phys.} {\bf #1} (#2) #3}
\def\lmp#1#2#3{{\it Lett. Math. Phys.} {\bf #1} (#2) #3}
\def\tmatp#1#2#3{{\it Theor. Math. Phys.} {\bf #1} (#2) #3}
\def\jhep#1#2#3{{\it JHEP} {\bf #1} (#2) #3}
\def\hepth#1{{\tt hep-th/}#1}
\font\zfont = cmss10 
\font\litfont = cmr6

\def\bigone{\hbox{1\kern -.23em {\rm l}}}     
\def\ZZ{\hbox{\zfont Z\kern-.4emZ}}
\def\hf{{\litfont {1 \over 2}}}

\def\p{\partial}

\def\b{\beta}

\def\m{\mu}

\def\s{\sigma}

\def\vp{\varphi}

\def\oo{\hat \omega   }

\def\o{\omega }


  \def\hb{\hbar\ }
  


  \def\CC {{\cal C}}
  \def\CD {{\cal D}}
  
  \def\CF {{\cal F}}

  \def\CO {{\cal O}}

  \def\CR {{\cal R}}
  \def\CS {{\cal S}}

  \def\CW {{\cal W}}

  \def\CZ {{\cal Z}}
 %

\chardef\tempcat=\the\catcode`\@ \catcode`\@=11
\def\cyracc{\def\u##1{\if \i##1\accent"24 i%
    \else \accent"24 ##1\fi }}
\newfam\cyrfam



\def\hb{\hbar}

  \def\R{\relax{\rm I\kern-.18em R}}
  \font\cmss=cmss10 \font\cmsss=cmss10 at 7pt 
  \def\Z{\relax\ifmmode\mathchoice
  {\hbox{\cmss Z\kern-.4em Z}}{\hbox{\cmss Z\kern-.4em Z}} 
  {\lower.9pt\hbox{\cmsss Z\kern-.4em Z}}
  {\lower1.2pt\hbox{\cmsss Z\kern-.4em Z}}\else{\cmss Z\kern-.4em 
  Z}\fi} 
  \def\p{\partial}
  
  \def\11{1\ \  1}
 \def\Det{{\rm Det }}

   \def\hepth{ {\tt hep-th/}}

 
\lref\HKK{ J. Hoppe, V. Kazakov, and I. Kostov,
``Dimensionally reduced SYM$_4$ as solvable matrix quantum
mechanics'', \np{571}{2000}{479},  \hepth{9907058}.}
\lref\NKK{V. Kazakov, I. Kostov and  N.  Nekrasov,
``D-particles, matrix integrals and KP hierachy'',
\np{557}{1999}{413},  \hepth{9810035}.}
\lref\KKK{V. Kazakov, I. Kostov, and D. Kutasov,
``A Matrix Model for the Two Dimensional Black Hole",
\np{622}{2002}{141}, \hepth{0101011}.}
\lref\AK{S. Alexandrov and V. Kazakov,
``Correlators in 2D string theory with vortex condensation'',
\np{610}{2001}{77}, \hepth{0104094}.}
\lref\IK{I. Kostov, ``String Equation for String Theory on a Circle'',
\np{624}{2002}{146}, \hepth{0107247}.}
\lref\DOUGLAS{M. R. Douglas,  ``Conformal theory techniques 
 in Large $N$ Yang-Mills Theory'', talk at the 1993
 Carg\`ese meeting, 
 \hepth{93}{11}{130}.}
\lref\KAZMIG{V. Kazakov and A. A. Migdal, \np{311}{1988}{171}.}
\lref\BRKA{E. Brezin, V. Kazakov and Al. Zamolodchikov,
\np{338}{1990}{673}.}
\lref\PARISI{ G. Parisi, \pl{238}{1990}{209, 213}.}
\lref\GRMI{ D. Gross and N. Miljkovic, \pl{238}{1990}{217}.}
\lref\GIZI{P. Ginsparg and J. Zinn-Justin, \pl{240}{1990}{333}.}
\lref\BIPZ{ E. Brezin, C. Itzykson, G. Parisi, and J.-B. Zuber,
\cmp{59}{1978}{35}.}
\lref\KPZDDK{KPZ, D, DK}
\lref\GRKL{D. Gross and I. Klebanov, \np{344}{1990}{475};
  \np{354}{1990}{459}.}
\lref\BULKA{D. Boulatov and V. Kazakov,  
``One-Dimensional String Theory with Vortices as Upside-Down
Matrix Oscillator'', \ijmp{8}{1993}{809}, \hepth{0012228}. }
 
\lref\DVV{R. Dijkgraaf, E. Verlinde, and H. Verlinde,
``String propagation in a black hole geometry'',
\np{371}{1992}{269}.}
\lref\FZZ{ V. Fateev, A. Zamolodchikov, and Al. Zamolodchikov,
{\it unpublished}.}
\lref\AKK{ S. Yu. Alexandrov, V. A. Kazakov, I. K. Kostov, 
``Time-dependent backgrounds of 2D string theory'',
\hepth{02}{05}{079}. }
\lref\MUKHIVAFA{S. Mukhi and C. Vafa,
``Two dimensional black-hole as a topological coset model of 
c=1 string theory'', \np{407}{1993}{667}, \hepth{9301083}.}
\lref\GOSHALVAFA{D. Ghoshal and C. Vafa,
``c=1 String as the Topological Theory of the Conifold'',
\np{453}{1995}{121}, \hepth{9506122}.}
\lref\KZ{V. A. Kazakov and A. Zeytlin,
``On free energy of 2-d black hole in bosonic string theory'', 
\jhep{0106}{2001{}021}, \hepth{0104138}.}
\lref\TECHNER{J. Teschner, 
``The deformed two-dimensional black hole'',
\pl{458}{1999}{257}, \hepth{9902189}. }
\lref\JAP{T. Fukuda and K. Hosomichi,
``Three-point Functions in Sine-Liouville Theory'',
\jhep{0109}{2001}{003}, \hepth{0105217}. }

\lref\MP{G. Moore and M. Plesser,  
``Classical scattering in 1+1 Dimensional string theory'',
\physrev{46}{1992}{1730}, \hepth{9203060}.} 
\lref\MPR{G. Moore, M. Plesser, and S. Ramgoolam,
``Exact S-matrix for 2D string theory'',
\np{377}{1992}{143}, \hepth{9111035}.}
\lref\MOORE{G. Moore, ``Gravitational phase transitions
  and the sine-Gordon model", \hepth{9203061}.}
\lref\DMP{R. Dijkgraaf, G. Moore, and M.R. Plesser,
``The partition function of 2d string theory'',
\np{394}{1993}{356}, \hepth{9208031}.}
\lref\sixauthors{E. Br\'ezin and V. Kazakov, \pl{236}{1990}{144};
M. Douglas and S. Shenker,  \np{335}{1990}{635};
D. Gross and A. Migdal, \prl{64}{1990}{127}.}

\lref\POLCHINSKI{J. Polchinski, ``What is string theory'',
{\it Lectures presented at the 1994 Les Houches Summer School
``Fluctuating Geometries in Statistical Mechanics and Field Theory''},  
\hepth{9411028}.}
\lref\KLEBANOV{I. Klebanov, {\it Lectures delivered at the ICTP
Spring School on String Theory and Quantum Gravity},
Trieste, April 1991, \hepth{9108019}.}
\lref\JEVICKI{A. Jevicki, ``Developments in 2D string theory'',
\hepth/9309115.}
\lref\HSU{E. Hsu and D. Kutasov, ``The Gravitational Sine-Gordon Model'', 
\np{396}{1993}{693}, \hepth{9212023}.}
\lref\MSS{G. Moore, N. Seiberg, and M. Staudacher,
``From loops to states in 2D quantum gravity'',
\np{362}{1991}{665}. }

\lref\JM{M. Jimbo and T. Miwa, ``Solitons and Infinite Dimensional
Lie Algebras'', {\it Publ. RIMS, Kyoto Univ.} {\bf 19}, No. 3
(1983) 943.}
\lref\Hir{R. Hirota, Direct Method in Soliton Theory {\it Solitons},
Ed. by R.K. Bullogh and R.J. Caudrey, Springer, 1980.}
\lref\UT{K. Ueno and K. Takasaki, ``Toda Lattice Hierarchy":
in `Group representations and systems of differential equations',
{\it Adv. Stud. Pure Math.} {\bf 4} (1984) 1.}
\lref\Takasak{K. Takasaki,
{\it Adv. Stud. Pure Math.} {\bf 4} (1984) 139.}
\lref\MukhiImbimbo{ C. Imbimbo and S. Mukhi,
``The topological matrix model of c=1 String", 
\np{449}{1995}{553}, \hepth{9505127}.}
\lref\NTT{ T.Nakatsu, K.Takasaki, and S.Tsujimaru, 
``Quantum and classical aspects of deformed $c=1$ strings'',
\np{443}{1995}{155}, \hepth{9501038}.}
\lref\Takebe{T. Takebe, ``Toda lattice hierarchy and conservation
 laws'',
\cmp{129}{1990}{129}.}
\lref\EK{T. Eguchi and H. Kanno,
``Toda lattice hierarchy and the topological description of the 
$c=1$ string theory'', \pl{331}{1994}{330}, hep-th/9404056.}
\lref\Krichever{I. Krichever, {\it Func. Anal. i ego pril.},
{\bf 22:3} (1988) 37  (English translation:
{\it Funct. Anal. Appl.} {\bf 22} (1989) 200);
``The  $\tau$-function of the
universal Witham hierarchy, matrix models and topological field
theories'', {\it Comm. Pure Appl. Math.} {\bf 47} (1992),
\hepth{9205110}.}
\lref\TakTakb{K.~Takasaki and T.~Takebe,
``Quasi-classical limit of Toda hierarchy and W-infinity symmetries'',
\lmp{28}{93}{165}, \hepth{9301070}.}
\lref\orlshu{A. Orlov and E. Shulman, \lmp{12}{1986}{171}.}
\lref\Nakatsu{T. Nakatsu, ``On the string equation at $c=1$'', 
\mpl{A9}{1994}{3313}, \hepth{9407096}.}
\lref\TakSE{K. Takasaki,
``Toda lattice hierarchy and generalized string equations'',
\cmp{181}{1996}{131}, \hepth{9506089}.}
\lref\TakTak{K. Takasaki and T. Takebe, 
``Integrable Hierarchies and Dispersionless Limit'',
\rmatp{7}{1995}{743}, \hepth{9405096}.}

\lref\kkvwz{ I. Kostov, I. Krichever, M. Mineev-Veinstein,
P. Wiegmann, and  A. Zabrodin, ``$\tau$-function
for analytic curves", \hepth{0005259}.}
\lref\Zabrodin{ A. Zabrodin, ``Dispersionless limit of Hirota
equations in some problems of complex analysis'',
\tmatp{129}{2001}{1511}; \tmatp{129}{2001}{239}, 
{\tt math.CV/0104169}.}
\lref\bmrwz{ A. Boyarsky, A. Marshakov,  O. Ruchhayskiy,
P. Wiegmann, and  A. Zabrodin, ``On associativity equations in
dispersionless integrable hierarchies", 
\pl{515}{2001}{483}, \hepth{0105260}.}
\lref\wz{P. Wiegmann and  A. Zabrodin, ``Conformal maps and
dispersionless integrable hierarchies", \cmp{213}{2000}{523},
\hepth{9909147}.}
\lref\FERTIGone{H.A. Fertig, \prb{36}{1987}{7969}. }
\lref\FERTIGtwo{H.A. Fertig, \prb{387}{1988}{996}. }
\lref\WIEGAGAM{O. Agam, E. Bettelheim, P. Wiegmann, and A. Zabrodin,
``Viscous fingering and a shape of an electronic droplet in 
the Quantum Hall regime'', {\tt cond-mat/0111333}.}
\lref\mwz{ M.~Mineev-Weinstein, P.~B.~Wiegmann, A.~Zabrodin, 
``Integrable Structure of Interface Dynamics'', \prl{84}{2000}{5106},
{\tt nlin.SI/0001007}.}
\lref\MVafa{ S.~Mukhi and C. Vafa, 
``Two-dimensional Black Hole as a Topological Coset
 Model of c = 1 String Theory", \hepth{9301083}.}
 
\lref\Faddeev{L.D. Faddeev, \lmp{34}{1995}{249}.}
 \lref\DFGZ{P. Di Francesco, P. Ginsparg and J. Zinn-Justin,
``2D gravity and random matrices'', 
{\it Phys. Rep.} {\bf 254} (1995). }
\lref\GGG{M. Gaudin, 
``Une famille \`a un param\`etre d'ensembles unitaires'',
{\it Nuclear Physics} {\bf 85} (1966) 545;
Travaux de Michel Gaudin, ``Mod\`eles exactement r\'esolus",
{\it Les Editions de Physique} 1995, p. 25.}
\lref\ADEM{I.K. Kostov,   ``Gauge invariant 
matrix models for the $\hat A-\hat D-\hat E$ closed strings", \pl{297}{1992}{74}, \hepth{9208053}.} 
\lref\KazMig{V.~Kazakov and A.~Migdal, \np {311}{171}{1989}. }
\lref\DougKdV{M. Douglas,
``Strings in less than one dimension and 
the generalized KdV hierarchies'',
\pl{238}{1990}{176}.}
 \lref\moore{G. Moore,  ``Double-scaled field theory at $c=1$",
\np{368}{1992}{557}.}
\lref\GinsMoore{P. Ginsparg and G. Moore, ``Lectures on 2-d gravity and 2-d string theory", \hepth{9304011}.}
\lref\WITTENGR{E. Witten, ``Ground Ring of
 two dimensional string theory'',
\np{373}{1992}{187}, \hepth{9108004}.}
\lref\MuchiImbimbo{C. Imbimbo
and S. Mukhi, ``The topological matrix model of c=1 String",
\hepth{9505127}.}
\lref\ddvfkn{R. Dijkgraaf, H. Verlinde, E.  Verlinde,
 ``Loop equations and Virasoro constraints in nonperturbative 2-d
quantum gravity", \np{348}{1991}{435};  M. Fukuma, H. Kawai and R. Nakayama, 
``Infinite dimensional grassmanian structure of the two-dimensional quantum gravity",
\cmp{143}{1992}{371}.}
 \lref\natspolch{M. Natsuume and J. Polchinski, ``Gravitationalscattering in the $c=1$
matrix model",
\np{424}{1994}{137}.}
 
\Title{\vbox{\baselineskip12pt\hbox
{}\hbox{ }}}
{\vbox{\centerline
 {Integrable Flows in }
 \centerline{$c=1$ String Theory}
\centerline{}}}
\vskip2pt
\centerline{Ivan K. Kostov\footnote{$^\circ$}{{\tt 
kostov@spht.saclay.cea.fr}}}
\bigskip

\centerline{{\it Max-Planck-Institut f\"ur  
Gravitationsphysik
}}
\centerline{{\it Albert-Einstein-Institut  
}}
\centerline{{\it Am Muehlenberg 1, D-14424 Golm, Germany
}}

\centerline{ \it and}

\centerline{{\it Service de Physique 
Th{\'e}orique, CEA/DSM/SPhT,}}
\centerline{{\it Unit\'e de recherche
 associ\'ee au CNRS,  CEA/Saclay}}
\centerline{{\it 91191
 Gif-sur-Yvette c\'edex, 
France\footnote{{\dag}}{Permanent address}}}

\vskip 1cm
\baselineskip8pt{

\baselineskip12pt{
\noindent

In these notes we review the
  method to construct integrable deformations
 of the compactified $c=1$  bosonic string theory
by primary fields (momentum or winding modes),
developed recently in collaboration 
with S. Alexandrov and V. Kazakov.
The method is based on the 
 formulation of the string theory as a
matrix model. The flows generated 
by either momentum or winding modes
(but not both) are  integrable
and satisfy the Toda lattice 
hierarchy\footnote{*}{Based on the talks 
presented at the James H. Simons
 Workshop on Random Matrix Theory
 Stony Brook, February 20 - 23, 2002, and the 
 International Conference on Theoretical Physics
 Paris, UNESCO, July 22-27, 2002}.
 }
\bigskip

\noindent

\Date{August, 2002}

\baselineskip=12pt 
\listtoc\writetoc 

\newsec{ Introduction}

The
$c=1$ string theory (the theory of random surfaces embedded in 1 dimension)
can be constructed as the collective theory for a 1-dimensional $N\times N$ hermitian matrix field \KLEBANOV. The  $U(N)$-invariant sector of the matrix
model is described by a non-relativistic quantum mechanics of free fermions in an 
upside-down quadratic potential.  The  mapping to free fermions allows to  calculate 
virtually any quantity in the string theory in all orders in the genus expansion. 

The elementary excitations of the $c=1$ string represent collective excitations of free fermions.  The tree-level
$S$-matrix can be extracted by considering the propagation of  ``pulses"
along the Fermi sea and their reflection off the ``Liouville wall"
\refs{\POLCHINSKI}.
 
The exact nonperturbative $S$-matrix   has been calculated 
by Moore {\it et al} \MPR . Each $S$-matrix element can be associated with a
single fermionic loop with a number of external lines.
One can then expect that the theory is also solvable in a nontrivial,
time-dependent background generated by a finite tachyonic source.
Dijkgraaf, Moore and Plesser \DMP\ demonstrated that this is indeed
the case when the allowed momenta form a lattice as in the case of the
compactified Euclidean theory.  In \DMP\ it has been shown that the
string theory compactified at any radius $R$ possesses the integrable
structure of the Toda lattice hierarchy \TakTak. 
 The operators associated with the  momentum modes  
in the string theory  have been interpreted in \DMP\ as Toda flows.   
 
The explicit construction of these  Toda 
flows    
 is an interesting and
 potentially important task,
because this would allow us to explore the 
 time-dependent  string backgrounds.
This problem was recently solved in \AKK.
The method used in \AKK\ is conceptually similar to the method of
orthogonal polynomials of Dyson-Mehta in the interpretation of 
M. Douglas \DougKdV , which has been used 
to solve the $c<1$ matrix models \refs{\DFGZ, \GinsMoore}.
The  construction of \AKK\ allows to evaluate the 
partition function of the string theory in 
presence of  finite perturbation by momentum modes.
The simplest case of such a perturbation is
the Sine-Liouville string theory\foot{It is conjectured 
 \refs{\FZZ} that  such a  perturbation 
  can lead to a target 
space geometry with a horizon. }.
   
Besides the momentum modes,  the compactified string  theory contains
 a second type of excitations associated with nontrivial windings 
around the target circle.  In the world-sheet description, the 
string theory represents a compactified gaussian field coupled
to  2d quantum gravity. Then the momentum and winding modes are
the electric and magnetic  operators  for this  gaussian field. 
It is known 
\refs{\GRKL, \BULKA}
that the winding modes  
propagate in the non-singlet 
sector of the $c=1$ matrix model,  which
is no more equivalent to a system of fermions.
Explicit realization of the winding modes 
can be given by gauging the 
compactified matrix model.
Then the winding modes 
are realized as  the Polyakov loops winding  around the
Euclidean time interval \KKK.

The electric and magnetic operators are exchanged by the 
usual electric-magnetic  duality  $R\to 1/R$ called also T-duality.
 If the momentum modes  of the compactified string theory 
are described by the Toda integrable flows, then by T-duality  the same 
is true also for the winding modes. 
A direct proof  of  is presented in \KKK, where it is  also shown that the
the grand-canonical partition function of the matrix model
perturbed by only winding modes  is a $\tau$-function of the 
Toda lattice hierarchy.  Applying  the T-duality backwards, we 
conclude that the same is true  for the partition function of the theory 
perturbed by  only  momentum modes.

A special case represents  the theory is compactified at the self-dual radius $R=1$,
it is equivalent to a topological theory that computes the Euler 
characteristic of the  moduli space of Riemann surfaces \MUKHIVAFA. 
 When $R=1$ and only in this case, the  partition function of the string theory 
has alternative realization as a Kontsevich-type model
\refs{\DMP,   
\MukhiImbimbo}.

  In this notes we summarize the results concerning the Toda 
integrable structure of the compactified $c=1$ string theory 
 obtained in refs. \refs{\KKK, \AK, \IK, \AKK}.
We will follow mostly the last work \AKK,
but will present the construction in the
framework of the 
 Euclidean 
compactified theory, 
while  \AKK\ discussed 
the theory in Minkowski 
space.   

\newsec{The matrix model for the $c=1$ string theory revisited}
\def\Xp{X_+ }
\def\Xm{X_- }
\def\Xpm{ X_{\pm}}

\subsec{Partition function of the 
Euclidean theory compactified on a circle with radius $R$}
The $c=1$  string theory 
(see Appendix A) describes the critical
behavior of  the large-$N$ matrix quantum mechanics
 (see the review \KLEBANOV\ 
and the references therein).
The critical point  is associated with the
maximum of the  matrix  potential. 
 In the scaling limit, which is dominated by
dense planar graphs,  the potential can be approximated
by a  quadratic one (with the wrong sign), 
which is stabilized by imposing a cut-off wall
far from the top.
The  relevant piece of the matrix Hamiltonian is
thus
\eqn\HamO{
H_0= \hf \Tr\ (P^2 - M^2), }
where $P=-i{\p/ \p M}$ and $M_{ij}$ is an
 $N\times N$ hermitian
matrix variable. The cosmological constant $\mu$ 
 is introduced as a ``chemical potential'' 
coupled to the size of the matrix $N$,
which should be considered as a dynamical variable.

We will consider the  Euclidean 
theory with periodic Euclidean
 time $x=-it$. The time interval
\eqn\betaR{\b=2\pi R}
can be also interpreted as 
inverse temperature.
The matrix model describes the $c=1$ string 
only when $\b>\b_{KT}= 4\pi$.
At $\b = \b_{KT} $ the 
winding   modes associated with the  $N(N-1)$
angular degrees of freedom  becomes important 
and produce a Berezinski-Kosterlitz-Thouless-like
phase transition to a $c=0$ string theory \GRKL.
The angular degrees of freedom can be interpreted as 
the winding modes of the string, which 
represent strings winding several times
 around the target circle . 

A sensible matrix model for  
 $\b<\b_{KT}$ can be constructed by 
introducing an additional 
  $SU(N)$ gauge field $A(x)$, which projects to
the sector free of winding modes \BULKA .
The  partition function of the 
compactified system is
\eqn\partF{\CZ(\mu)=\sum_{N\ge 0} e^{-\b N} \CZ_N,}
where 
$\CZ_N$ is given by a functional integral
with respect to the one-dimensional matrix 
hermitian fields 
$M(x)$ and $A(x)$ with periodic boundary 
conditions 
\eqn\partFI{
\CZ _N =  
\int\limits_{\rm periodic  } 
\CD P  \CD M\CD A\  e^{-{1\over \hbar}S_0}.
}
The action  functional is
\eqn\fractMP{
 S_0= \Tr\  
\int ^{ \b} _0   \left( iP\nabla_A M 
 - H_0\right)
 dx,
}
 $\nabla_A M$ denotes the covariant time derivative
\eqn\timeD{\nabla_A M = \p_x M -i [A, M].}
The gauge field can be also used to study perturbations
by  winding modes, with appropriately 
tuned coupling constants, so that the long range order
is not destroyed. 
It was shown in \KKK\ that the winding modes 
couple to the moments of the $SU(N)$ magnetic
flux associated with the gauge field $A$.

\subsec{Chiral quantization}

The analysis of the matrix model simplifies
considerably if it is formulated in
terms of the   chiral
 variables \AKK

\eqn\YpYm{\Xpm = {M\pm P\over \sqrt{2}}}
representing $N\times N $ hermitian matrices.
The  Hamiltonian 
in the new variables is

 \eqn\hammat{
 H_0 = -  \hf \Tr (\hat  \Xp\hat\Xm+\hat \Xm\hat\Xp),}
 where the matrix 
operators $\hat\Xpm$
obey the canonical commutation relation

\eqn\ccrYY{ [(\hat\Xp)^i_j,(\hat\Xm)^k_l]=-i
  \ \delta^{i}_{l}\delta^k_j.}
The partition function 
for fixed $N$ 
is now given
 by a path  integral 
with respect to the one-dimensional hermitian matrix fields
$\Xp(x)$, $\Xm(x)$ and $A(x)$,
satisfying  periodic boundary conditions. The 
action \fractMP\ reads, in terms of the new fields,
\eqn\GENFZ{S_0=
 \Tr\  
\int ^{\b} _0   \left(i \Xp 
\nabla _A \Xm -\hf  \Xp\Xm\right)
 dx.}
 The partition function \GENFZ\ depends on
 the gauge field only 
 through the global holonomy factor,
given by the unitary matrix
\eqn\defOMEG{\Omega =
 \hat T e^{i\int_0^{2\pi R} A(x)dx}.
 }
This can be seen by imposing the gauge $A=0$,
 which can be done at the expense of 
replacing the periodic boundary condition 
$\Xpm(\b)=\Xpm(0)$ by 
a $SU(N)$-twisted one
\eqn\twibc{ \Xp(\b) = \Xp(0), \qquad \Xm(\b)
=
 \Omega^{-1}
 \Xm(0)\ \Omega.}

\subsec{Gauge-invariant collective excitations:
momentum and winding modes}

 The  momentum modes $V_{n/R}$ of the 
matrix field are gauge-invariant
  operators 
 with time evolution $V_{n/R}(t) = e^{-n t/R}V_{n/R}(0)$. 
 Here $n$ can be any positive or negative integer.
 The spectrum of momenta is determined by
 the condition that the time 
 dependence is periodic 
in the imaginary direction with period $2\pi R$.  
  The operators satisfying these conditions   are
\eqn\vertopM{  V_{n/R}=
 \ e^{\pm i nx/R}\  \Tr\Xpm(x)^{n/R} = 
 \cases{\Tr \Xp^{|n|/R} & if $n>0$\cr
\Tr \Xm^{|n|/R} & if $n<0$}.}
This form of the left and right  momentum modes 
 was first suggested to our knowledge by
Jevicki \JEVICKI.

As it was shown in \KKK, the  winding modes $\tilde V_{nR}$  are
 associated with the $SU(N)$ magnetic 
flux through the compactified spacetime 
\eqn\vortopM{ \tilde V_{nR} = \Tr\  \Omega^n}
where $\Omega$ is the $U(N)$ holonomy \defOMEG.
 From the world sheet  point of view the 
 winding operator 
 $\tilde V_{nR}$  creates  a  puncture with
 a Kosterlitz-Thouless vortex, $i.e.$ 
  a line of discontinuity $2\pi Rn$ 
 starting at  the puncture.

\subsec{The partition function  as a three-matrix integral }

In the $A=0$ gauge the path integral  with respect to the
fields $\Xpm(x)$  is gaussian with  the determinant 
of the quadratic form equal to one.
Therefore it is reduced to the integral with respect to the 
initial values $\Xpm=\Xpm(0)$
of the  action \GENFZ\ evaluated 
along the classical trajectories, which satisfy the
twisted periodic boundary condition \twibc.
Therefore  the canonical partition function of the
matrix model can be 
reformulated as an ordinary matrix integral
with respect to the two hermitian matrices 
 $\Xp$ and $\Xm$, 
 and the unitary matrix
$\Omega$:
    \eqn\partvr{
\CZ_N=
 \int  D \Xp
D \Xm
  d\Omega \  
\ e^{ i\Tr (\Xp\Xm -  q \Xm \Omega \Xp \Omega^{-1} ) },
}
 where we denoted 
\eqn\qqqq{q= e^{i\b}= e^{2\pi i R}.}

A general  perturbations by momentum 
modes can be  introduced 
by changing the homogeneous measures
 $D\Xpm$ to
    \eqn\deformmes{
[D\Xpm] = D\Xpm \ e^{\pm i\Tr  U_\pm(\Xpm)},
}
where we introduced the matrix potentials   
\eqn\verpotpm{
  U_\pm (\Xpm) =R  \sum_{n>0}  t_{\pm n}\Tr 
  \Xpm^{ n/R }  }
Similarly, a general
perturbation  by winding 
modes
can be  introduced 
by changing the invariant  measure
 $D\Omega$ on the group $U(N)$ 
 to
    \eqn\deformOM{
[D\Omega] = D\Omega \ e^{\tilde U(q^{1/2} 
\Omega)- \tilde U(q^{-1/2} \Omega)}}
where we introduced the 
following matrix potential   
%
%
 \eqn\vorpot{\tilde U(\Omega) =
    \sum_{n\ne 0} \tilde t_n  
 \Tr\  \Omega^{n}.}

In case of a generic perturbation with 
both momentum and winding
 modes, the three-matrix model \partvr\ 
is not integrable. 
However, it can be solved exactly in 
case of an arbitrary perturbation 
only by  momentum or only by  winding modes.
In the following we will discuss in details
 these two integrable cases.
In both cases it is possible to integrate
 with respect to the angles and 
the matrix partition function reduces to
 an eigenvalue integral.

 \subsec{Integration with respect to the angles}

\def\xm{ x_- }
\def\xp{ x_+}
\def\xpm{ x_\pm  }
 
Let us  consider the case of a perturbation with 
only momentum modes. Then the gauge
 field plays the role of a 
Lagrange multiplier for the condition $[\Xp,\Xm]=0$
and the two matrices can be simultaneously diagonalized.
Applying twice the Harish-Chandra-Itzykson-Zuber
 formula 
 we can reduce the matrix integral to
 an integral over the
eigenvalues $x^\pm_1,\dots, x^\pm_N$ of 
the two hermitian matrices
 matrices $\Xpm$.  We
write, after rescaling the integration variables,
\eqn\partrvd{\CZ_N(t)=\int\limits _{-\infty}^\infty
 \prod_{k=1}^N
 [d x^+_k ][d x^-_k] \ 
\det_{jk}\left( e^{i x^-_j x^+_{k}}\right)\ 
\det_{jk} \left( e^{-iq x^-_{j} x^+_k}\right)}
where the measures are defined by
\eqn\measuresxpm{[d\xpm]=d \xpm \  e^{i U_\pm(\xpm)}. 
}
Then the  grand canonical partition function 
can be written as a Fredholm determinant
\eqn\grpf{  Z(\mu, t)=\Det(1+e^{-\b \mu} K _+ K_- )}
where  
\eqn\evolhf{\eqalign{
[K_+ f](\xm) &=
  \int  [d\xp]\ 
e^{  i \xp \xm} f(\xp) \cr
[K_- f](\xp) &= \int
 [d \xm]\  e^{ - iq\xp \xm} f(\xm) .
}}

 %

\newsec{The partition function of the 
non-perturbed theory}

 \subsec{Eigenfunctions of the non-perturbed 
Hamiltonian.}

In the absence of perturbation, the partition function
is given by the Fredholm determinant
 \eqn\partfHo{ \CZ(\mu) = \Det(1+ e^{-\b(\mu+H_0)})}
and can be interpreted as the grand canonical
finite-temperature partition function  for
a system of non-interacting fermions 
in the inverse gaussian potential.
The Fredholm determinant \partfHo\
can be computed once we know 
a complete set of eigenfunctions for the
  one-particle   Hamiltonian 
\eqn\oneph{H_0
= -\hf (\hat   \xp\hat \xm+\hat \xm\hat \xp).}
The canonically conjugated operators 
 $\hat\xp$ and  $\hat\xm$
\eqn\ygrekpm{
 \hat \xp \hat \xm- \hat \xm \hat \xp    =-i}
are represented  in the space of functions of $\xp$ as
$$  \hat \xp=\xp, \ \ \hat \xm = i\p_{\xp},$$
and in the space of functions of $\xm$ 
 as
$$  \hat \xm=\xm, \ \ \hat \xp= -i\p_{\xm}.$$
%

\def\Ezpm{ \left\langle  E | \xpm\right\rangle}

%

Since the inverse gaussian potential is bottomless,
the spectrum of the  Hamiltonian 
is  continuous 
\eqn\schro{ H_0
 \psi^E_+(\xpm)\equiv \mp i (\xpm\p_{\xpm} +1/2)\psi^E_+(\xpm)
=
 E\psi^E_+(\xpm) \qquad (E\in \R).}
The eigenfunctions in the $\xpm$-representation, 
which we denote by Dirac brackets
$$\psi^E_\pm(\xpm) = \Ezpm $$ 
are given by 
\eqn\wavefp{
\left\langle  E | \xpm\right\rangle
= {1\over \sqrt{\pi}}\ e^{\mp \hf i\phi_0}\ 
\xp^{\pm i E-\hf}}
where the phase factor $\phi_0=\phi_0(E)$ 
will be determined in a moment\foot{The eigenfunctions 
\wavefp\ play in the 
chiral quantization the same role as  the parabolic cylinder 
functions in the standard quantization, based on the
original Hamiltonian \HamO.}.
 The solutions \wavefp\   have a
 branch point at  $\xpm=0$ and 
are defined unambiguously only on the positive 
real axis.
For each energy $E$ there are 
two solutions, since
 there are two ways to define the 
analytic continuation to the negative axis. 
The wave functions relevant for our problem
 have large negative energies and 
are supported, up to  exponentially small
 terms due to tunnelling   phenomena, 
 by the positive axis\foot{This corresponds to the 
theory of type I of \MPR.}.
Therefore we define the analytic continuation
  to   negative axis as
\eqn\extNA{\eqalign{ \left\langle  E |
 -\!\xp\right\rangle&=
 \left\langle  E | e^{-i\pi}
 \xp\right\rangle=e^{\pi E}
 \left\langle  E |  \xp\right\rangle
\qquad (\xp>0)\cr
\left\langle E |-\!\xm \right\rangle
&=\left\langle  E |e^{i\pi}\xm \right\rangle
=e^{\pi E}\left\langle E|  \xm  \right\rangle
\qquad(\xm>0).}}
 The functions \wavefp\ with 
$E$ real form two 
orthonormal sets
\eqn\normEE{
 \int\limits_{-\infty}^\infty d \xpm
\langle E | \xpm\rangle\langle\xpm| E' \rangle=
\delta(E-E').}
We will  impose 
also  the bi-orthogonality condition
\eqn\biortcn{ 
 \int\!\!\!\!\!\!\int\limits_{-\infty}^\infty
 {d \xp d\xm\over \sqrt{2\pi}}
 \langle E | \xp\rangle \ e^{ i \xp\xm}
\langle\xm | E' \rangle= 
\delta(E-E'),}
which fixes the phase $\phi_0$ 
\eqn\rfacts{e^{i\phi_0(E)}
  =  \sqrt{1\over 2 \pi}
e^{-{\pi\over 2} (E- i/2)}  \Gamma( iE + 1/2). 
} 
The phase defined by \rfacts\ has
in fact  a small
 imaginary part, which we neglect\foot{The 
pure phase factor would be 
$  \sqrt{\cosh \pi E \over 2\pi}
 \Gamma( iE + 1/2)= \sqrt{1+e^{\pi E}} e^{i\phi_0(E)}$.
}.
 This is because the theory restricted to the Hilbert
 space spanned on these
  states is   strictly speaking not unitary
due to the tunneling phenomena across the top of
 the potential.
In the following we will systematically
neglect the   exponentially small 
 terms $\CO(e^{\pi E})$.
With this accuracy we can write the
completeness condition
\eqn\compltns{
 \int\limits_{-\infty}^\infty dE 
 \langle \xp |E\rangle
 \langle E| \xm\rangle
= {1\over \sqrt{2\pi}}\ e^{-
i\xp\xm}  ,}
where the right-hand side is the kernel of the
inverse Fourier transformation relating the
$\xp$ and $\xm$ representations. 
 For real energy the phase $\phi_0$ is real
 (up to non-perturbative terms),
but  we will analytically continue it to
 the whole complex plane. In this case 
the reality condition means
\eqn\unitrtybr{ \overline{\phi_0( E) }
 =  \phi_0(\bar E).}

 \subsec{Cut-off prescription, density
 of states and free energy}

 To find the density of states, we need to introduce
  a cutoff $\Lambda$
such that $\Lambda>> \mu$. This can be done by 
putting a completely reflecting wall at distance 
${\xp+\xm\over \sqrt{2}}=\sqrt{2\Lambda}$ from the origin.
Since the wall is completely reflecting, there is no flow of  momentum through it,
${\xp -\xm\over \sqrt{2}}=0$.
Thus  the cutoff wall is equivalent to the following boundary condition 
at $\xp=\xm=\sqrt{\Lambda}$
 \eqn\bcctff{ \psi ^E_+(\sqrt{\Lambda}) = 
 \psi ^E_-(\sqrt{\Lambda})}
on the wavefunctions
 \wavefp. This condition  is satisfied for a discrete 
set of energies $E_n\ \ (\ n\in \Z )$
defined by
 \eqn\kvconbis{\phi_0(E_n) - E_n\log \Lambda 
 +2\pi n=0.
 }
From  \kvconbis\ we can find the density
 of the energy levels in the 
confined system
 \eqn\DEN{
   \rho(E)=     {\log \Lambda\over 2 \pi}-
 {1\over 2\pi } {d\phi_0(E)  \over d E} }
%
Now we can calculate free energy
 $\CF(\mu, R)= \log\CZ(\mu, R)$ as
 \eqn\FRENO{
 \CF(\mu, R)=  \int_{-\infty}^\infty
 d E\, \rho(E)\log\left[1+e^{-\beta(\mu+E)}\right]
 }
with the density \DEN. Integrating  by parts 
in \FRENO\ and dropping out the 
$\Lambda$-dependent piece,
we get
 \eqn\FRENOo{
 \CF(\mu, R) =-{1\over 2\pi}\int d\phi_0 (E)
 \log \left( 1+e^{-\beta(\mu+ E)}\right)=-
R \int_{-\infty}^{\infty}dE
 { \phi_0(E) \over 1+e^{\beta(\mu+ E)}} .
 }
 We close the contour of
 integration in the upper half plane
and  take the integral as a sum of residues.
This gives for the free energy
 \eqn\FRENOoo{\CF = -i \sum _{r=n+\hf > 0}
 \phi_0\left( i r/R-\mu\right).}
Using the explicit form \rfacts\ of   $\phi_0$ 
 we can represent  the free energy  as  a sum over pairs 
 of positive half-integers
 \eqn\frenergy{  
\CF(\mu, R)   =\sum_{r, s\ge 1/2}^{\infty}
 \log\left(\mu + i r+ {is/  R}   \right)
}
which is explicitly invariant under the T-duality 
$R\to 1/R$, $\mu\to R\mu$.
From here it follows that the free energy satisfies
 the  functional equation
\eqn\funceQ{ 4  \sin \left( \p_\mu/2R\right)
 \sin \left( \p_\mu/2\right)\CF(\mu, R)=-\log\mu  .}
 From \FRENOoo\ we can express the phase factor 
in terms of the free energy
\eqn\phiF{ \phi _0
=- 2 { \sin \left( \p_\mu/2R\right)}\CF(\mu, R).}

\subsec{Calculation of the free energy
using the $E$-representation of the operators $\hat\xpm$}

Let us now give an alternative, algebraic  derivation of 
  the free energy  of the non-perturbed theory.
For this purpose we will  write the canonical commutation 
relation \ygrekpm\ for the matrix elements of 
the operators $\hat \xpm$  in  $E$-representation. 

The action of the operators  $\hat \xpm$ 
to the wave functions 
is equivalent to  a shift of the energy
 by an imaginary unit
(we consider the wave functions as
 analytic functions of $E$) 
and a multiplication by a phase factor:
\eqn\psiHzz{ \eqalign{
  \langle E|  \hat \xpm |    \xpm\rangle  & = 
 e^{\pm \hf i[\phi_0(E\mp i)-\phi_0(E)]}
  \  \langle E\mp i|\xpm\rangle 
}}
As a consequence, the operators $\hat\xpm$
 are represented by
the finite difference  operators 
 acting in the $E$-space
\eqn\bareREP{ 
\hat\xpm \ \ \rightarrow \ \ e^{\mp \hf i\phi}
 e^{\mp i\p_E}e^{\pm \hf i\phi}.
}
The  Heisenberg relation \ygrekpm\ 
is equivalent to the condition
\eqn\StreQ{\phi_0(E+\hf i)-\phi_0(E-\hf i) = - \log (-E) 
}
which  is equivalent
 to the functional
constraint \funceQ.

\newsec{Integrable perturbations
by momentum modes.}

\subsec{One-particle eigenfunctions and  
the density of states of deformed system}

Now we will consider  perturbations 
generated by momentum modes $V_{n/R}$.
As we argued  in sect. 2.4, this can be achieved
by    deforming  the 
integration measures $d\xpm$ to
\eqn\measureD{[d\xpm]=
 d\xpm  \exp\left(\pm iU_\pm(\xpm)\right), 
\quad U_\pm (\xpm) 
= R\sum\limits_{k\ge 1} t_{\pm k} \xpm^{k/R} . }
   In this section we will show that such a
 deformation is exactly solvable,
being generated by a system of commuting flows
$H_n$ associated with the coupling constants $t_{\pm n}$.
The associated integrable structure is that
 of  a constrained Toda Lattice
hierarchy. The method
is very similar to the standard Lax formalism 
of Toda theory, 
but we will not assume that the reader is familiar with
this subject. It is based on  the possibility 
to describe the 
perturbation by vertex operators as deformations of the 
$E$-representation  of the canonical
commutation relation.

The  partition function of the perturbed  system is 
given by the 
Fredholm determinant \grpf, where  the integration
 kernels \grpf\ and \evolhf\ are defined with the 
deformed measures \measureD. 
The deformed kernel is  diagonalized by a 
complete orthonormal system of wave functions\foot{As 
before, we
 understand the
completeness in a weak sense, $i.e.$ up to 
non-perturbative
terms due to tunnelings phenomena, which we 
 have neglected.}
$\Psi_\pm^E (\xpm)$
\eqn\ortnRM{ \int d\xpm \overline{\Psi_\pm^E
 (\xpm)} \Psi_\pm^E (\xpm)=
\delta(E-E')}
which
satisfy the following three conditions:

\noindent
1)  they are eigenfunctions of the 
evolution operator 
relating the points  $x=0$ and $x=\b$
\eqn\eigHo{ e^{-\b H_0} \Psi^E_\pm = e^{-\b E}
 \Psi^E_\pm,}

\noindent
2) they behave at infinity as
\eqn\asymptPsi{\Psi^E_\pm (\xpm)
 \sim \xpm ^{\pm iE-\hf}\  e^{ \mp \hf i \phi(E) }\
e^{i U_\pm (\xpm)},}

\noindent
3)
 they satisfy the  bi-orthogonality condition  
\biortcn
\eqn\ortonorm{
 \int\!\!\!\!\!\int
 {d \xp d\xm\over \sqrt{2\pi}}\ \
 \overline{ \Psi^E_- (\xm)}\ 
  \ e^{ i \xp\xm} \Psi_+^{E'} (\xp)
= 
\delta(E-E').}

 The new eigenfunctions 
are  not necessarily eigenvalues of the Hamiltonian
$H_0$ itself. Indeed, the evolution operator $e^{-\b H_0}$
acts trivially on any entire
 function of $\xpm^{1/R}$.
In particular, the condition \asymptPsi\  
 is compatible 
with the condition \eigHo.

Once the basis of the perturbed wave functions is found,
the logarithm of the Fredholm determinant can be 
calculated as the integral \FRENO, where the 
new density of states is obtained from the
 phase $\phi(E)$ 
in the asymptotics \asymptPsi\ 
\eqn\rhoPer{\rho(E) = - {1\over 2\pi } {d\phi\over dE}.}
Therefore the phase $\phi(E)$ contains 
all the information about the
perturbed system. It is related to the free energy
by 
\eqn\FRENOoo{\CF = -i \sum _{r\ge 1/2} \phi
 \left( i r/R-\mu\right).}
  \eqn\phiF{ \phi (-\mu)
= 2 { \sin \left(\hb  \p_\mu/2R\right)}\CF(\mu, R).}

\subsec{Dressing operators}

Now we proceed to the  actual calculation
of the phase $\phi$.
The method is a generalization of the 
algebraic method we have used in sect. 3.3.
First we remark that  the  perturbed
wave functions are given by  the rhs of \asymptPsi\
up to  factors of the form 
\eqn\Wfactrs{\eqalign{
W_\pm(\xpm) 
= & \exp\left(iR \sum_{n\ge 1} 
v_{\pm n}\  \xpm^{-n/R}\right).
}}
which satisfy the second condition \eigHo\ and
tend to $1$ when $\xpm\to\infty$.
The unknown coefficients
$v_n$) and the phase $\phi$
are determined   functions of $E$ and 
the couplings $t_n$ by  the third condition 
\ortonorm.

It follows from \bareREP\ that the 
deformed wave function
\eqn\Drss{\Psi^E_\pm (\xpm) =e^{i U_\pm (\xpm)}
e^{\mp\hf  i \phi(E)} \xpm^{\pm i E-\hf}
 W_\pm(\xpm)
}
can be obtained from the
bare  wave functions \wavefp\
by acting 
with two finite-difference operators in the $E$-space
 $\hat \CW_+$ and $\hat \CW_-$
\eqn\PertWf{  \Psi_\pm^E (\xpm)= 
 \langle E|e^{\pm\hf i\phi_0}\hat \CW_\pm   | \xpm\rangle.}
The explicit form of the  dressing operators
$\hat\CW_\pm$ is
obtained by replacing $\xpm\to e^{\mp i \p_E}$
in \Drss. The dressing operators 
are unitary 
 \eqn\isometry{\hat
  \CW_+ \hat \CW_+^{\dag} = \CW_-\hat \CW_-^{\dag} = 
1}
since they relate two
 orthonormal systems of functions. 
Further, the bi-orthogonality condition  \ortonorm\ 
is equivalent to the   identity
\eqn\StrW{\hat  \CW_-^{\dag}  e^{i\phi_0}\hat
 \CW_+ =   1.}
   The identity
\StrW\ means that the product
$\hat\CW_-\hat\CW_+^{-1}$ 
does not depend on the perturbation, which is 
a general property of all Toda lattice systems
\TakTak.

 \subsec{Lax  operators and string equation}

We have seen that the partition function is expressed
 in terms of the phase $\phi(-\mu)$.
Therefore we 
assume that the energy is at the Fermi level $E=-\mu$,
 and  consider 
all variables as functions of $\mu$ instead of $E$. 
Let us denote by $\oo$ the operator 
\eqn\defom{ \oo = e^{i\p_\m}}
shifting the variable $\m$ by $i$.
The operators $\oo$ and $\mu$ 
satisfy the Heisenberg-Weyl commutation relation
\eqn\HWcomr{ [\oo, \mu]= \oo, \qquad
[\oo^{-1}, \mu]= - \oo^{-1}.}

Now let us consider the representation of these 
commutation relations in the perturbed theory.
The  dressing operators $\CW_\pm$ 
are now   exponents of 
series in $\oo$ with $\m$-dependent coefficients
\eqn\WWpm{\hat \CW_\pm = e^{ iR
\sum_{n\ge 1} t_{\pm n} \oo^{n/R}}\ 
e^{\mp\hf  i\phi(\mu)}\ 
e^{ iR
\sum_{n\ge 1} v_{\pm n}(\mu)\  \oo^{-n/R}}.
}
The operators
\eqn\LaxO{\eqalign{
L_+ &= \CW_+ \oo 
   \CW^{-1}_+, \qquad L_- = \CW_- \oo ^{-1}
   \CW^{-1}_-,\cr
M_+ &= \CW_+ \mu 
   \CW^{-1}_+, \qquad M_- = \CW_- \m 
   \CW^{-1}_-.
}
}
known as Lax and Orlov-Schulman operators
satisfy the same commutation 
relations  as the operators $\oo$ and $\mu$
\eqn\LaxLM{
[L_+, M_+] =  i L_+, \qquad[L_-, M_-] = - i L_-
.}
The Lax operators $L_\pm$ 
 represent 
the canonical coordinates $\hat \xpm$
in the basis of perturbed wave functions,
\eqn\LaxPpm{\langle E |e^{\pm\hf i\phi_0}\hat \CW_\pm  
 L_\pm |\xpm\rangle =\langle E |e^{\pm\hf i\phi_0}\hat \CW_\pm  \hat\xpm
 |\xpm\rangle
}
 while 
the Orlov-Shulman operators $M_\pm$ 
represent
Hamiltonian $H_0 = -
\hf(\hat \xp\hat\xm+\hat\xm\hat\xp)$. 
Therefore the $L$ and $M$  operators 
 are related also by
\eqn\STREq{ M_+= M_- = \hf(L_+L_-+L_-L_+).}
The last identity  is  not 
satisfied automatically in the Toda system 
and represent an additional constraint 
analogous to the string equations 
in the minimal models of 2D quantum gravity.
The relation \STREq\ is proved by 
inserting the identity \StrW\ in the 
bare relation for  $E=-\mu$,
  satisfied by the operators \bareREP.
 The string equation can be 
also written as the Heisenberg commutation relation 
between the two Lax operators
\eqn\ComLpm{ [L_+, L_-]= -i}
which also follows directly from  \ygrekpm.

The operators $M_\pm$ can be 
expanded as infinite series of the 
$L$-operators. Indeed, as they act to the
dressed wave functions as 
\eqn\OrlSPepm{\eqalign{&\langle E |\;
e^{\pm\hf i\phi_0}\; \hat \CW_\pm \; 
 M_\pm |\xpm\rangle
= \pm i (\xpm \p_{\xpm} -1/2)
\Psi^E_\pm (\xpm)\cr
 &= \left(\sum _{k\ge 1} 
 k \ t_{\pm k} \ \xpm^{ k/R}  +\m+
\sum_{k\ge 1} v_{\pm k} \
 \xpm ^{-k/R}\right)\Psi^E_\pm (\xpm ).
}
}
we can write
\eqn\OrlSh{
M_\pm =
   \sum _{k\ge 1}  k t_{\pm k}   L_\pm^{ k/R}+\mu  +
\sum_{k\ge 1} v_{\pm k}  L_\pm^{-k/R}.
}

In order to exploit the Lax equations \LaxLM\
and the string equations \STREq\
we need the explicit form of the two Lax operators.
It follows from  \WWpm\ 
that $L_\pm$
can be represented as series of the form
\eqn\LpLmG{ \eqalign{
L_+ &=  e^{- i\phi/2}
\left(\oo   +\sum_{k\ge 1} a_{ k}\
   \oo^{ 1- n/R} \right)e^{ i \phi/2}
\cr
 L_- &=  e^{ i\phi/2}
\left( \oo^{-1}  +\sum_{k\ge 1} a_{- k} 
 \ \oo^{- 1+ n/R}
 \right)e^{- i \phi/2}.
}
}
%

\subsec{Integrable flows} 

Let us identify the integrable flows associated
with the coupling constants $t_n$.
From the 
definition \LaxO\ we have
\eqn\TodA{ 
\p_{t_n} L_\pm  = [ H_n, L_\pm], }
where the operators $H_n$ are related to the
 dressing operators as
\eqn\HWpm{
H_{n} = (\p_{t_ {n}}\CW_+) \CW_+^{-1} =
 (\p_{t_{n}}\CW_-) \CW_-^{-1}.}
The two representations of the flows
 $H_n$ are equivalent by virtue of
the relations \isometry\ and \StrW. 
 A more explicit expression in terms of the
 Lax operators
is derived by the following standard argument. 
 Let us consider the
case $n>0$. From the explicit form of the
 dressing operators
it is clear that $H_n = W_+
\oo^{n/R} W_+^{-1} + $ negative 
powers of $\oo^{1/R}$. The variation
of $t_n$ will change only the coefficients 
of the expansions \LpLmG \
of the Lax operators, preserving their 
general form.  But it is clear
that if the expansion of $H_n$ contained 
negative powers of
$\oo^{1/R}$, its commutator with $L_-$ would 
create extra powers
$\oo^{-1 - k/R}$.  Therefore
\eqn\hashka{
 H_{ \pm n} =  (L_\pm^{n/R}   ) _{^{>}_{<} }
 +\hf (L_\pm^{n/R}   ) _{0} , 
\qquad n>0,
}
where the symbol $( \ \  )_{^{>}_{<}}$ means the positive
(negative) parts of the series in the shift operator
$ \oo ^{1/R}$ and $( \ \ )_{ 0}$ means
 the constant part. 
 By a similar argument one shows that the 
Lax equations \TodA\
are equivalent to the zero-curvature conditions
\eqn\zerocur{ \p_{t_m} H_n -\p_{t_n}H_m - [H_m, H_n]=0.}

Equations \TodA\ and  \TodA\ imply that
the perturbed theory possesses the Toda lattice
integrable structure.
The Toda structure implies  an 
infinite hierarchy  of PDE's for the
coefficients $v_n$ of the dressing operators, 
the first of which is
the Toda equation for the phase $\phi(\mu)
\equiv \phi(E=-\mu)$
\eqn\Todaeq{i{\p \over \p t_1}
{\p \over \p t_{-1}} \phi(\mu)
= e^{i\phi(\mu)-i\phi(\mu-i/R)}
 - e^{i\phi(\mu+i/R)-i\phi(\mu)} .  }
 The uniqueness of the solution is
 assured by appropriate boundary
conditions \KKK, which are equivalent to the 
 constraint \ComLpm.

\subsec{Representation in terms of a  bosonic field}

The momentum modes can be described as the
oscillator modes of a bosonic field $\vp(\xp, \xm)
= \vp_+(\xp)+ \vp_-(\xm)$.
The bosonization formula is 
\eqn\boseF{
\Psi^{E=-\mu-i/2} _\pm (\xpm)= \CZ^{-1}\ 
e^{\pm  i \vp_\pm (\xpm)}\cdot  \CZ}
where $\CZ$ is the partition function and 
\eqn\bosvp{
\vp_\pm(\xpm) =  
+R \sum _{k\ge 1} t_k \xpm^{k/R} +
 {1\over R} \p_\m  +
\mu\log \xpm -R \sum_ {k\ge 1} 
{1\over k} \xpm^{-k/R}{\p\over \p t_k}.
}
Then by \OrlSPepm\ the operators $M_\pm$ are represented by
the currents $\xpm \p_\pm\vp$ 
\eqn\crrnts{
M_\pm^{\dag} \Psi _\pm ^{E}(\xpm)|_{E=-\mu-i/2} 
=  \CZ^{-1} \  \xpm \p_\pm\vp \cdot  \CZ.}

 \subsec{The dispersionless (quasiclassical) limit}

Let us reintroduce the Planck constant
 by replacing $\mu \to \mu/ \hb$
and consider the quasiclassical limit $\hb\to 0$.
 In this limit the integrable
structure described above reduces to the 
dispersionless Toda hierarchy
\refs{\Krichever, \TakTakb, \TakTak}, 
where the operators $\mu$ and
$\oo$ can be considered as a pair of 
classical canonical variables
with Poisson bracket
\eqn\ccrom{ \{  \o, \mu  \} = \o. }
Similarly, all operators become $c$-functions 
of these variables.
The Lax operators can be identified
with the classical phase space 
coordinates $\xpm $, which 
satisfy
\eqn\canXpm{ \{ \xp, \xm\}=1.}
The two functions
  $\xpm(\o, E) $ define the 
classical phase-space trajectories 
as functions of the proper time variable 
$\tau = \log \o.$

The shape of the Fermi sea is 
determined by the  classical trajectory 
corresponding to the Fermi level $E=-\mu$.
In the non-perturbed system 
the classical trajectory  is 
\eqn\paramFT{
\xp(\o) = \sqrt{\mu}\ \o, \qquad
\xm(\o) = \sqrt{\mu}\ \o^{-1}}
and the Fermi sea has a hyperbolic shape
\eqn\FreFS{ \xp\xm = \mu.}

In the perturbed theory the
classical trajectories are of the form
\eqn\clastrP{\xpm = L_\pm (\o, \mu)}
where the functions $L_\pm$ are of the form 
\eqn\zoom{
L_\pm (\o, \mu)= e^{\hf \p_\mu\phi }\ 
\o^{\pm 1}
\left(1+\sum_{k\ge 1} a_{\pm k}(\mu)\ 
\o^{\mp k/R}\right).
} 
The flows $H_n$ become  Hamiltonians
for the evolution along the `times'
$t_n$. The unitary operators $\CW_\pm$
becomes a pair of canonical transformations 
between the variables $\o, \mu$ and
$L_\pm, M_\pm$. Their generating functions are 
given by the 
 expectation values 
$S_\pm = \CZ^{-1}\  \vp_\pm(\xpm)\ \cdot 
\CZ$ of the
chiral components of the bosonic field $\phi$
\eqn\EIcon{
S_\pm = \pm R\sum _{k\ge 1}   t_{\pm  k} \ 
 \xpm^{ k/R}  +\mu \log \xpm -\hf  \phi  \pm
R\sum_{k\ge 1} {1\over k} 
 v_k\   \xpm ^{-k/R}
}
where $v_k = \p \CF/\p t_k$.
The differential of the function $S_\pm$ is
 \eqn\eikonal{
dS_\pm = M_\pm d\log \xpm + \log \o \ d\mu +
R \sum_{n\ne 0} H_n dt_n.}
If we consider 
the coordinate $\o$ as a function 
of either $\xp$ or $\xm$, then
\eqn\omgxpm{
\o = e^{\p_\m S_+(\xp)} =e^{\p_\m S_-(\xm)}.}

The classical string equation
\eqn\conteq{\xp\xm =M_+ = M_-
} 
can be written, using  the expansion \OrlSh\
of $M_\pm$, as  
\eqn\TODApr{\eqalign{
\xp\xm 
&=\sum _{k\ge 1}  k t_{ k} \ 
 \xp^{ k/R}  +\mu  +
\sum_{k\ge 1} v_{ k}\  \xp ^{-k/R}\cr
\xp\xm 
&=\sum _{k\ge 1}  k t_{-k} \ 
 \xm ^{ k/R}  +\mu  +
\sum_{k\ge 1} v_{- k}\  \xm ^{-k/R}.
\cr}}
The first of these expansions is convergent  for
 sufficiently large $\xp$ and the second one
 for sufficiently large $\xm$.
Comparing  the two equations one can extract the 
form of the Fermi surface. Technically this is done
as follows \IK. 
First, note that if all $t_{\pm k}$ with $k>n$ vanish,
the sum in \zoom\ can be restricted to $k\le n$.
Then it is enough to substitute the 
expressions \zoom\ in the
profile equations \TODApr\ and compare 
the coefficients in front of 
$\o^{\pm k/R}$. 

 The two expansions \TODApr\ can be combined into 
a single equation 
\eqn\Hpmn{\xm\xm  - \sum _{k\ne 0}k  t_k H_k (\o)
   = \mu}
where 
\eqn\hamka{
H_{\pm k} (\o)= [  L_\pm ^{ k/R}(\o)]_{^>_<}
+\hf [  L_+^{ k/R}(\o)]_0\qquad (k>0).}
The left hand side can be interpreted  as the Hamiltonian
for the perturbed system. 
It defines the 
profile  of the perturbed Fermi sea, 
which is a deformation of the 
hyperbole \FreFS.

\subsec{One-point correlators
in the dispersionless limit}
It follows from  $\log \o= \p_\mu S_\pm(\xpm)$ that 
the $\mu$-derivative of  the one-point correlators
\eqn\vevv{\langle \Tr \Xpm^{n/R}\rangle=  {\p\CF  \over \p t_{n}},
}
is given by the contour integrals 
\eqn\opcoro{
  {\p^2\CF  \over\p\mu \ \p t_{\pm n}}={1\over 2\pi i}
 \oint {d \o^{1/R} \  [L_\pm (\o)]^{|n|/R}
 } \qquad (n\ge 1)}
where the closed contour of integration in the variable
 $\o^{1/R}$  goes along the
arc between $\o = e^{-i\pi R}$ 
and $\o = e^{i\pi R}$. (Note that 
the integrand is expanded in  Laurent series 
in $\o^{1/R}$.)

\subsec{Example: sine-Gordon field 
coupled to 2D gravity}

 The simplest nontrivial string theory 
with time-dependent background
is the sine-Gordon theory coupled to 
gravity known also as
Sine-Liouville theory. It is obtained 
by perturbing with the
lowest couplings $t_1$ and $t_{-1}$. 
In
this case 
 \eqn\zomSG{
\xpm = e^{\hf\p_\mu \phi } 
 \o^{\pm 1} (1+ a_{\pm 1} \o ^{\mp {1\over R}} ).}
and \TODApr\ give
\eqn\coefzom{\eqalign{
&\mu\  e^{-\p_\mu\phi } - 
\left(1-{1\over R}\right)t_1 t_{-1} 
e^{-(2-{1\over R}) \p_\mu\phi} =1,
\cr 
& a_{\pm 1} =  t_{\mp 1}  \ e^{-\hf (2- {1\over R})
\p_\mu\phi}.  } }
The first equation is an
algebraic equation for the
 the susceptibility
$$u_0= \p_\mu^2\CF = - R \p_\mu \phi.$$
It 
 was first found (for the T-dual theory)
 in  \KKK. This algebraic equation resumes
the perturbative expansion found 
in  \MOORE.
The equation \zomSG\ with $a_\pm$ given in \coefzom\
was first found in \AK\ 
by integrating the Hirota equations for 
the Toda hierarchy with the boundary 
condition given by the non-perturbed free energy
\frenergy.

\newsec{Integrable perturbations by 
winding modes}

\subsec{The partition function 
as
integral over the $U(N)$ group}

Let us consider a perturbation of the $c=1$
string theory by only winding modes.
The relevant variable in this case is the gauge field $A(x)$
and the matrix model \partFI/ can be viewed as  a
two-dimensional gauge theory defined on the disc
having as a boundary the compactified Euclidean time
interval. The only nontrivial degree of freedom
in such a theory is  the holonomy factor around the circle
\defOMEG. The canonical partition function 
is given by the $U(N)$-integral
\eqn\partfUN{
\CZ_N= \int {[D\Omega]\over
|q^{-1/2} \ \Omega\otimes I- q^{1/2} \ I\times  \Omega|}
}
where $I$ is the unit $N\times N$ matrix.
The integrand depends only on the 
eigenvalues $z_1, ..., z_N$ of 
the unitary matrix $\Omega$ 
and the partition function becomes
\eqn\zeden{\CZ_N(\tilde t_n )=  |q^{1/2 }- q^{-1/2}|^{-N}
 \oint\prod_{k=1}^N \oint {[dz_k] 
\over 
 2\pi  i  z_k}\
 \prod_{ j< k}\left|{ z_j-z_k\over 
q^{-1/2}  z_j - 
 q^{1/2 } z_k }\right|^2 .
 }
where the integration goes along the unit circle $|z|=1$
with measure
\eqn\measureZ{ [dz]= dz  \ e^{\tilde U(q^{\hf} z)
-\tilde U(q^{-\hf} z) }.}
This representation of the partition function
was studied (for the non-perturbed theory)
by Bulatov and Kazakov \BULKA. 
Note that the absolute value can be abandoned, since 
the integrand is homogeneous.

   The  
 grand canonical partition function 
   can be written as the absolute value of 
   a Fredholm determinant
 \eqn\Zdet{\CZ(\mu,  t_n)= 
     \left|{\rm Det} (1+ q^{i\mu}  \hat K)\right|,}
    where the Fredholm kernel 
is defined by the contour integral
    \eqn\Kkernel{(\hat Kf)(z)=
 - \oint {[dz]\over 2\pi i}\ 
    {f(z')
    \over q^{1/2} z- q^{-1/2} z'}.
    }
 The integral has poles when $z_i=qz_j$ 
and should be evaluated by 
adopting a prescription
for surrounding the poles.
The prescription used in \BULKA\ is
to add a small imaginary part to $R$
so that  $|q|<1$. In the case of real $q\in[0,1)$
the partition function was studied by M. Gaudin \GGG.

\subsec{Evaluation of the partition function of the 
non-perturbed theory}

Boulatov and Kazakov showed in \BULKA\ that
 the prescription
for the contour integration 
gives for the Fredholm determinant \Zdet\ 
the same result as \frenergy, up to non-perturbative terms
$\CO(e^{-\pi\mu})$. 
Here we recall their calculation.
The integration kernel \Kkernel\ 
acts to the monomials $z^n$ as
\eqn\kernZ{ K z^n = \cases{ q^{n+\hf} z^n & if $n\ge 0$\cr
0
& if  $ n<0$}}
and  the Fredholm determinant  \Zdet\ reads 
 $$ \CZ(\mu) = \prod_{r>0} (1+ q^{i\mu +r}).
 $$
 The free energy 
then is equal to an infinite sum
\eqn\DualFrgy{\CF(\mu)=
\sum_{n\ge 0} \log(1+q^{i\mu +n+1/2})
=
\sum_{m\ge 1} { (-)^m q^{im\m}\over m}
 {1\over q^{m/2}-q^{-m/2}}}

The last sum is the result 
can be written as the integral
 \eqn\intrepFR{\CF(\mu)  
=
\int\limits _{\CC} {dy\over y} \ {q^{i\mu y}
 \over 
4 \sinh (  yR)
\  \sinh (   y)}.}
along a contour $\CC$ circling around the poles $y= in, \ n>0$ only.
If $|q|<1$, then $\CC$ 
goes from $-\infty$ to $0$ and then up the imaginary axis
to $i\infty$.
 
Note that if we close the contour around the
 poles of $\sinh \pi Ry$, then
 the result will be the sum over the residues $k= inR \ (n>0)$ 
which can be written as 
the partition function 
for the dual radius $\tilde R= 1/R$.

\subsec{Toda integrable structure}

The Fredholm determinant  \Zdet\ with 
non-homogeneous measure \measureZ\  
can be represented as the Fock expectation 
value in a theory of chiral fermions defined on 
the unit circle \KKK.
The deformations by winding modes are introduced as
Bogolyubov transformations of the left and right fermionic
vacua and the partition function was identified
as  a $\tau$-function 
of the Toda Lattice hierarchy. 

As a consequence, the free energy satisfies an
infinite hierarchy  of PDE with the Toda 
 `times' $\tilde t_n$.
The first one is the Toda equation dual to the
equation \Todaeq\ 
\eqn\TodaDu{
i{\p\over \p\tilde t_1}{\p\over \p\tilde t_{-1}}
\tilde \phi (\mu)  =
e^{i\tilde \phi(\mu) - i\tilde \phi(\mu-i)}
- e^{i\tilde \phi(\mu+i) - i\tilde \phi(\mu)}
}
where
\eqn\tildephi{\tilde \phi(\mu)
= 2 \sin (\p_\mu/2) \CF(\mu, R).}
Using the scaling relation given in the Appendix,
one can reduce \TodaDu\ to an ordinary differential
 equation, which should be solved with initial condition 
\intrepFR.
In the quasi-classical (genus zero) limit,
this differential equation can be integrated to
an algebraic equation for the string 
susceptibility $u_0 = \p_\mu \tilde\phi$
\eqn\suscar{\mu e^{{1\over R} \p_\mu\tilde\phi }+
 \ \tilde t_1 \tilde t_{-1} (R-1)
e^{{2-R\over R} \p_\mu\tilde\phi}  =1.}
The one- and two-point correlators were 
calculated by Alexandrov and Kazakov
from the higher equations of the
Toda hierarchy \AK.
These results were later 
confirmed  in \IK\ using the 
 Lax formalizm of the constrained Toda system.

\subsec{Lax operators and string equation}

We will 
consider  the potential
$\tilde U(z)$ in the measure \measureZ\ 
as the value on the unit circle $\bar z z=1$
of the  potential 
\eqn\Utiolde{ \tilde U(z,\bar z)=
\sum _{n\ge 0} (\tilde t_n z^n + \tilde t_{-n} \bar z^n)}
defined in the whole complex plane. 
Let us assume  that the 
spectral  variables $z$ and $\bar z$ 
are represented (in some sense)
 by a pair of Lax operators
of the form
\eqn\Laxdual{\eqalign{
L&=e^{- \hf i \tilde\phi}\ \oo  \left( 1+ \sum _{k\ge 1} 
u_k \oo ^{1-k}\right)e^{ \hf i \tilde\phi}\cr
\bar L&=e^{ \hf i \tilde\phi}\ \oo^{-1}  \left( 1+ \sum _{k\ge 1} 
u_k \oo ^{1-k}\right)e^{- \hf i \tilde\phi}
,}}
where the phase $\tilde \phi$ is related to the 
free energy by
\eqn\tildePhi{\tilde\phi (\mu)
= -i[\CF(\mu+i/2) - \CF(\mu -i/2)].}

In order to find the constraint satisfied by $L$ and $\bar L$,
we consider the non-perturbed theory, for which the expressions 
in the parentheses are equal to one.
The functional equation \funceQ\ 
means that the bare phase $\tilde \phi_0$ satisfies 
\eqn\funcQphi{  \tilde \phi_0(\mu +i/R) -  \tilde \phi_0(\mu -i/R)
= -i\log\mu.}
In the case of no perturbation, this is equivalent to the 
algebraic relations 
\eqn\strQDu{
L^{1/R} \bar L^{1/R}+\bar L^{1/R} L^{1/R}
= \mu}
and
\eqn\strqHDu{ [L^{1/R}, \bar L^{1/R}]= i.}
The second identity is  invariant with respect to 
the dressing procedure and therefore are satisfied by the
the general Lax operators \Laxdual.
This is the  string equation for the constrained Toda system.

\subsec{Quasiclassical limit
 and relation to the conformal map problem}

In the dispersionless limit $\mu\to\infty$
the two Lax operators define a smooth closed curve $\gamma$
in the complex plane, whose equation is written
 in a parametric form as
\eqn\curveG{ z = L(\o),\quad  \bar z=\bar L(\o)
\qquad (|\o|=1).}
For example, in the case when only  $\tilde t_1$ 
and $\tilde t_{-1}$
are nonzero, the explicit form of the curve $\gamma$
is
\eqn\crfc{\eqalign{
 z&=  e^{\hf \p_\m \tilde\phi }\ \o \ (1+ \tilde t_{-1}
 e^{{2-R\over 2R}\p_\m \tilde\phi} \ \o^{-1})^R\cr
 \bar z &=  e^{-\hf \p_\m \tilde\phi}\ \o^{-1}\ (1+ \tilde t_1
 e^{{2-R\over 2R}\p_\m \tilde\phi} \ \o)^R.}
 } 

Assuming that the couplings $\tilde t_n$ and $\tilde  t_{-n} $ are 
complex conjugate,
the map \curveG\
can be extended to a conformal map
from the exterior of the unit disk $\o<1$ to
a connected  domain $\CD$ 
 with the topology of a disk
containing the point $z=\infty$ and 
bounded by the curve $\gamma$.
 The couplings $\tilde t_n \ ( n\ne 0)$ and $\mu$ 
 can be thought of as a set 
coordinates in the space of closed curves. 
The relation between the conformal maps and the 
dispersionless Toda hierarchy was studied recently 
in a series of papers
 \refs{\wz, 
 \kkvwz,  \Zabrodin}.

\newsec{Conclusion}

In these notes we explained how the 
 the 
perturbations of the  
compactified $c=1$ 
string theory  by momentum or winding modes 
are described by integrable deformations of a 
gauged matrix model on a circle.
The momentum and winding modes are associated 
with the collective excitations of the
matter and gauge fields.
The deformed system is described by 
a constrained 
Toda Lattice hierarchy.
The Lax formalism for  the   Toda lattice hierarchy allows to 
calculate explicitly the free energy and the correlation functions
of the electric or magnetic operators 
for any genus. In particular, the partition function
is a $\tau$-function of the Toda lattice hierarchy.

The integrability takes place only in the grand canonical 
ensemble, in which  the size $N$ is a dynamical variable
and the string interaction constant is controlled by  
the  chemical potential $\mu$.
The partition function is a Fredholm determinant, 
and not a usual determinant as in the case of the  open 
 matrix chains for   the $c<1$ string theories.

It is not likely that the integrability is conserved for 
perturbations with {both} moment and winding modes.
Nevertheless, the calculation of the 
correlation functions of the momentum modes in presence of
a perturbation modes (or vice-versa) seem to
be performable albeit very difficult. 
This calculation might help to check the 
hypothesis (related to the FZZ conjecture\FZZ)
that a strong perturbation by winding modes
can lead to  a curved background
 with Euclidean horizon \KKK .

Another  problem, also related to the 
\FZZ\ conjecture, is to find the 
limit $\mu/t_1t_{-1}\to 0$ of the matrix model,
 which  is relevant to  the 
sine-Liouville string theory. 
This limit 
seems to be subtle
because at  small  $\mu $ the 
non-perturbative
effects can enter into the game.
In any case, the $\mu\to 0$ limit of the correlators 
of the matrix model 
does not seem to reproduce  the results 
 obtained in the Sine-Liouville theory \refs{\FZZ, \JAP}.

Finally, it would be very interesting to
understand the origin of the integrability
from the point of view of the 
world-sheet string theory.

     \smallskip\smallskip\smallskip
 \bigskip
\noindent
{\bf  Acknowledgments}
\smallskip

\noindent
We thank the Albert Enstein Institut, Golm, for hospitality.
This research is supported in part by European network 
EUROGRID HPRN-CT-1999-00161.

\appendix{A}{Momentum and winding modes in
 the  $c=1$ Euclidean string theory}

In this section we will recall briefly the
 world-sheet description of the $c=1$
   Euclidean string theory with compact 
target space.  
The elementary excitations in this theory
 are  closed surfaces, or string world sheets,  
embedded in a circle of radius $R$.    
An embedded  surface is  defined by  metric 
 tensor $g_{ab}(\s)$ $(a,b=1,2)$
and the position $x(\s)$ in the target space 
 as functions of the local coordinates 
$\s=\{\s_1, \s_2\}$.  The free energy of the 
string theory 
is given by the functional integral  over all
 connected surfaces
\eqn\frenE{{1\over \hb^2} 
\CF(R, \mu, \hb )=
\int \CD g_{ab}\CD x\ e^{-\CS(g_{ab}, x)}}
with a weight  given by the   the Polyakov 
 action
\eqn\PSTR{
\CS(g_{ab} , x)={1\over 4\pi}\int \limits_{\rm world\ 
 sheet} d^2\sigma\sqrt{{\rm det} g}
[ g^{ab}\p_a x\p_b x +4\pi \mu +  \hat \CR^{(2)} 
 \log \hb],  }
where $\hat \CR $ is the local curvature 
associated with the metric $g_{ab}$.
The parameter $\mu$ is the cosmological
 constant, coupled to the area of the
world sheet, and $\hb$ is the string 
coupling constant, which 
 is associated with   the processes  of splitting
 and  joining of closed strings.
 By the Euler theorem the global curvature is
 related to the genus $h$ of the 
world sheet as
\eqn\glcur{{1\over 4\pi}\int d^2
\sigma\sqrt{{\rm det} g}\  \hat \CR^{(2)}= 2-2h
}
 and one can write the free energy  
 as a series  
 \eqn\genexp{ \CF(R,g) =\sum_{h\ge 0}
  \hb^{2h} \CF^{(h)}(R,\mu).}

 In the conformal gauge  
$g_{ab}=e^{- 2\phi(\s)}\delta_{ab}$,  the 
  conformal factor  
$\phi$ becomes   a dynamical field due to 
the conformal anomaly, and
 the world-sheet action  becomes essentially 
a $c=1$ conformal field theory
coupled to a Liouville field 
\eqn\confstr{
\CS={1\over4\pi}\int \limits_{\rm world\  sheet}
 d^2 \sigma\, [(\partial x)^2
+(\partial\phi)^2 +  \mu e^{b\phi} +\hat
 \CR^{(2)} (\log \hb +Q\phi)  +{\rm ghosts}].}
 The background charge $Q$ and the exponent
 $b$ of
 the Liouville field  are determined by  the 
requirement that the total 
conformal anomaly  vanishes
\eqn\ccge{c_{\rm matter} +c_{\rm Liouville} + 
c_{\rm ghosts} = 
1 + (1+ 6Q^2) - 26 =0}
 and that the perturbation due 
to the cosmological term $\mu e^{b\phi}$ is marginal 
\eqn\bexpli{ {b(2Q-b)\over 4 }=1.}
These two conditions give 
  \eqn\Qandb{Q=-2, \ \ b = -2.}
(The value $Q=2$ does not lead to a sensible
 classical limit.)
The invariance of the  action  with respect to 
shifts $\phi \to \phi +\phi_0$ implies that the 
free energy depends on 
$\mu$ and $\hb$ through the dimensionless 
combination $\mu \hb$, which is the statement
 of the double scaling limit in the $c=1$ 
string theory. 
Therefore we are free to choose $\hb=1$ and write
 the genus expansion \genexp\ as 
an expansion in $1/\mu^2$.

The primary operators associated with the matter  field  $x(\s)$ are 
the vertex operators  $V_e(\s) \sim e^{-i e x(\s)}$ and the
 Kosterlitz-Thouless
 vortices $\tilde V_m(\s)$. The operator $V_m(\s)$ is associated with 
 a   discontinuity $2\pi m$ of the field $x$
around  the point $\s$ on the world sheet.
We call   $e$ and $m$ electric 
and magnetic charges, in analogy 
with the Coulomb gas on the plane.  
The electric and magnetic charges should satisfy 
the Dirac condition $em =2\pi\times $integer.
From the point of view of the compactified
 string theory observables, the electric and 
magnetic charges are the   momentum and winding 
numbers, correspondingly.
The spectrum of the electric and magnetic charges
 for a periodic target space 
$x+2\pi R\equiv x$  is  
\eqn\spem{ e = n/R, \ m= nR \quad (n \in \Z).}

If we write the position field $x$ as a sum of 
a holomorphic and antiholomorphic parts,
$x = x_R+ x_L$, then the     Kosterlitz-Thouless 
vortices   are described by the
vertex operators for the 
 dual  field  $\tilde x = x_R- x_L$, whose target
 space is the circle of radius $1/R$. 
 When integrated 
over the world sheet, these operators should be accompanied
 by nontrivial Liouville factors compensating their 
anomalous dimensions. 
The  integrated operators have the form
\eqn\vvintegr{\eqalign{
 V_{e}
& \sim
\int d^2 \sigma\,  e^{-i  e x}e^{(e-2)\phi}, \cr
 \tilde V_{m}  &
\sim
\int d^2 \sigma\,  e^{-i  m \tilde x  }e^{(m-2)\phi}. }}
The Liouville exponents are determined by  
the condition that the
integrands  are densities. 
 
  We are interested in deformations of the
 string theory 
obtained by allowing   electric  charges
 $e=n/R$ with fugacities $t_n$ and
magnetic charges $m= nR$ with fugacities
 $\tilde t_n$.
This is achieved by adding to the action
 \confstr\ the perturbation term
\eqn\PERTS{ \delta\CS = \sum_{n\ne 0}( t_n V_{n/R} 
+\tilde t_n\tilde V_{nR}).  } 

 The translational invariance of the 
functional measure $\CD\phi$ yields the following 
Word identity for the free energy
\eqn\scalinglaw{-2 \hb {\p\CF \over \p \hb }-2 \m{\p\CF \over \p\m}
+\sum_{n\ne 0}  \left( nR-2\right)   t_n{\p\CF \over \p t_n} 
+ \sum_{n\ne 0} \left( {n\over R}-2\right)  \tilde t_n{ \p\CF \over \p \tilde t_n}= 0.}
 This  means that the  couplings  $t_n$,
 $\tilde t_n$ and the string coupling $\hb$
scale  with respect to the cosmological
coupling $\mu$ 
\eqn\FSCAL{ 
t_n\sim \mu^{1-\hf|n|/R},\quad
\tilde t_n\sim\mu^{1-\hf R|n|},\quad   
\hb\sim\mu^{-1}.  }
The  T-duality symmetry 
of the  original theory  
\eqn\dualsy{ x\leftrightarrow \tilde x, \ \ \ R\to1/R,\ \ \ 
 \mu\to \mu/R}
 holds for the perturbed theory if one also 
exchanges the couplings as
$t_n\leftrightarrow\tilde t_n$ 
(up to a rescaling by an $R$-dependent factor).

In the perturbation \PERTS\ we  should 
retain only the relevant  charges
$|e|< 2/R$ and $|m|<2R$. However, the correlation 
functions of a finite number of 
 irrelevant operators   are perfectly  meaningful.
The interesting phase of the deformed theory, 
which can be thought of as Coulomb gas  coupled to 
quantum gravity, is the plasma phase, where the 
fugacities of the 
charges are tuned so that the Debye length is of the 
order of the  size of the 2D universe.  

 %


\listrefs
\bye